\documentclass[11pt]{article}
\usepackage{graphicx}
\usepackage[margin=1.25in]{geometry}
\usepackage[usenames,dvipsnames]{color}
\usepackage{url}
\usepackage[colorlinks = true,
            linkcolor = blue,
            urlcolor  = blue,
            citecolor = blue,
            anchorcolor = blue]{hyperref}

\usepackage[numbers,compress]{natbib}

\newenvironment{Abstract}{\begin{quotation} \begin{center}
                       ABSTRACT
     \end{center}\bigskip  }{\end{quotation}}

\def\Acknowledgements{\bigskip  \bigskip \begin{center} \begin{large}
             \bf ACKNOWLEDGEMENTS \end{large}\end{center}}

\def\Acknowledgements{\bigskip  \bigskip \begin{center} \begin{large}
             \bf ACKNOWLEDGEMENTS \end{large}\end{center}}

\def\beq{\begin{equation}}
\def\eeq#1{\label{#1}\end{equation}}
\def\eeqn{\end{equation}}

\newenvironment{Eqnarray}%
   {\arraycolsep 0.14em\begin{eqnarray}}{\end{eqnarray}}
\def\beqa{\begin{Eqnarray}}
\def\eeqa#1{\label{#1}\end{Eqnarray}}
\def\eeqan{\end{Eqnarray}}

\let\bar=\overbar

\def\bra#1{\left\langle{ #1} \right|}

\def\lsim{\mathrel{\raise.3ex\hbox{$<$\kern-.75em\lower1ex\hbox{$\sim$}}}}
\def\gsim{\mathrel{\raise.3ex\hbox{$>$\kern-.75em\lower1ex\hbox{$\sim$}}}}

\def\del{\partial}
\def\Dslash{\not{\hbox{\kern-4pt $D$}}}
\def\dslash{\not{\hbox{\kern-2pt $\del$}}}
\def\pslash{\not{\hbox{\kern-2pt $p$}}}
\def\ETmiss{\not{\hbox{\kern-4pt $E$}}_T}

\def\Dlr{\mathrel{\raise1.5ex\hbox{$\leftrightarrow$\kern-1em\lower1.5ex\hbox{$D$}}}}

\def\MSB{{\bar{M \kern -2pt S}}}
\def\msb{{\bar{\scriptsize M \kern -1pt S}}}

\def\drb{{\bar{\scriptsize D \kern -1pt R}}}

\newcommand\snowmass{\begin{center}\rule[-0.2in]{\hsize}{0.01in}\\\rule{\hsize}{0.01in}\\
\vskip 0.1in Submitted to the  Proceedings of the US Community Study\\ 
on the Future of Particle Physics (Snowmass 2021)\\ 
\rule{\hsize}{0.01in}\\\rule[+0.2in]{\hsize}{0.01in} \end{center}}

\usepackage[utf8x]{inputenc}
\usepackage{amsmath}
\usepackage{amsfonts}
\usepackage{amssymb}

\def\prn#1{{\left(#1\right)}}

\def\abrk#1{{\langle#1\rangle}}

\def\bra#1{{\langle#1|}}

\def\cg(#1,#2)(#3,#4)(#5,#6){\bra{#1,#2,#3,#4}#5,#6\rangle}

\def\ts#1{{_{\mbox{\scriptsize #1}}}}

\def\threej(#1,#2)(#3,#4)(#5,#6){\begin{pmatrix}#1&#3&#5\\#2&#4&#6\end{pmatrix}}
\def\sixj(#1,#2,#3)(#4,#5,#6){\begin{Bmatrix}#1&#2&#3\\#4&#5&#6\end{Bmatrix}}
\def\ninej(#1,#2,#3)(#4,#5,#6)(#7,#8,#9){\begin{Bmatrix}#1&#2&#3\\#4&#5&#6\\#7&#8&#9\end{Bmatrix}}

\def\bs{\boldsymbol}
\def\mc{\mathcal}

\usepackage[arrows]{ezedits} 

\defineEdit{DK}{\color[rgb]{.7,.9,.8}}{\color[rgb]{0,.7,.3}}
\defineEdit{SK}{\color{purple}}{\color[rgb]{.26, .0, .33}}
\defineEdit{SR}{\color{blue}}{\color[rgb]{0, .20, .8}}
\defineEdit{AS}{\color{red}}{\color[rgb]{0.80, .20, .0}}

\usepackage[firstpage=true]{background}
\backgroundsetup{contents={\parbox{6.5in}{\snowmass}}, scale=1,placement=top,opacity=1,color=black,position={3.25in,1.2in}}

\usepackage{fancyhdr}
\fancypagestyle{plain}{%
  \fancyhf{}%
  \fancyhead[C]{}
  \fancyfoot[C]{\thepage}
}

\fancypagestyle{empty}{%
  \fancyhf{}%
  \fancyhead[C]{{\it Snowmass 2021 IF1 Quantum Spin-based Sensors White Paper}}
  \fancyfoot[C]{\thepage}
}
\title{Quantum sensors for high precision measurements of spin-dependent interactions}
\date{}

\usepackage{authblk}

\author[1,2]{Dmitry~Budker}
\author[3]{Thomas Cecil}
\author[4]{Timothy E. Chupp}
\author[5]{Andrew~A.~Geraci}
\author[6]{Derek~F.~Jackson~Kimball}
\author[7]{Shimon~Kolkowitz}
\author[8]{Surjeet~Rajendran}
\author[9]{Jaideep~T.~Singh}
\author[10]{Alexander~O.~Sushkov}

\affil[1]{Helmholtz-Institut and Johannes Gutenberg-Universit\"at, 55128 Mainz, Germany}
\affil[2]{Department of Physics, University of California at Berkeley, Berkeley, California 94720-7300, USA}
\affil[3]{Argonne National Laboratory, Lemont, IL 60439, USA}
\affil[4]{Department of Physics, University of Michigan, Ann Arbor, Michigan 48109, USA}
\affil[5]{Center for Fundamental Physics, Northwestern University, Evanston, IL 60208, USA}
\affil[6]{Department of Physics, California State University -- East Bay, Hayward, California 94542, USA}
\affil[7]{Department of Physics, University of Wisconsin, Madison, Wisconsin 53706, USA}
\affil[8]{Department of Physics and Astronomy, The Johns Hopkins University, Baltimore, Maryland 21218, USA}
\affil[8]{Facility for Rare Isotope Beams, Michigan State University, East Lansing, MI, 48824, USA}
\affil[10]{Department of Physics, Boston University, Boston, Massachusetts 02215, USA}

\begin{document}


\maketitle

\bigskip

\medskip

\begin{Abstract}
\noindent The applications of spin-based quantum sensors to measurements probing fundamental physics are surveyed. 
Experimental methods and technologies developed for quantum information science have rapidly advanced in recent years, and these tools enable increasingly precise control and measurement of spin dynamics.
Theories of beyond-the-Standard-Model physics predict, for example, symmetry violating electromagnetic moments aligned with particle spins, exotic spin-dependent forces, coupling of spins to ultralight bosonic dark matter fields, and changes to the local environment that affect spins.
Spin-based quantum sensors can be used to search for these myriad phenomena, and offer a methodology for tests of fundamental physics that is complementary to particle colliders and large scale particle detectors.
Areas of technological development that can significantly enhance the sensitivity of spin-based quantum sensors to new physics are highlighted.
\end{Abstract}


\def\thefootnote{\fnsymbol{footnote}}
\setcounter{footnote}{0}
%

\section{Executive Summary}
\label{sec:intro}

There are disparate profound mysteries in fundamental physics, ranging from the nature of dark matter and dark energy to the origin of the matter-antimatter asymmetry of the universe, and, in turn, a plethora of theoretical proposals to explain these mysteries. Unfortunately there are currently few if any clear experimental signatures indicating how best to unravel these mysteries. 
Consequently, in this era it is advantageous to cast a wide net in the search for new physics. 
A powerful, versatile, and relatively low-cost approach is to use the techniques, systems, and devices developed in the rapidly-growing field of quantum information science (QIS). 
Quantum systems can be made extremely sensitive to external perturbations. 
Indeed, much of the work in quantum science is focused on how to minimize this sensitivity, in order to prevent decoherence. 
Here we outline a complementary approach, which seeks to optimize the sensitivity of quantum systems to new fundamental physics.

There is a growing number of experiments that make use of quantum resources and systems to search for spin-dependent interactions of novel origin, which are predicted by a wide variety of beyond-the-Standard-Model physics theories~\cite{demille2017probing,safronova2018search}.
Experimental techniques for precision measurement of such spin-dependent interactions have substantially advanced over recent decades, in no small part because they share a common foundation with the robust program of research on spin-based quantum sensors for measurement of magnetic fields, magnetic resonance phenomena, and related phenomena. 
Furthermore, control and measurement of spins, spin ensembles, and quantum materials is at the heart of QIS and quantum computing schemes \cite{demille2002quantum,koch2019quantum,henriet2020quantum}. 
Thus the development of spin-based quantum sensors offers an opportunity for cross-fertilization between fundamental and applied research. 

In the context of searches for beyond-the-Standard-Model physics, precision measurements using the tools of QIS, magnetic resonance, and atomic, molecular, and optical (AMO) physics are complementary to collider-based high energy physics research. 
Precision experiments searching for discrete-symmetry-violating permanent electric dipole moments (EDMs), exotic spin-dependent interactions mediated by new light bosons, and spin-dependent couplings to ultralight bosonic dark matter fields [e.g., axions, axionlike particles (ALPs), and dark/hidden photons] can probe new physics associated with energy scales far beyond that capable with modern particle colliders \cite{demille2017probing,safronova2018search}.
This is because precision measurement experiments are designed to detect extremely subtle energy shifts (at scales $\sim 10^{-26}~{\rm eV}$). 
Because of their energy resolution, such experiments can be sensitive to physics generated by new high-mass particles.
For example, EDM searches are now sensitive to CP-violation due to virtual particles with masses $M \gtrsim 10~{\rm TeV}/c^2$.
Precision magnetic resonance-based search for axion-like dark matter are sensitive to axion-like particles arising from spontaneous symmetry breaking at scales $f_a$ reaching up to the GUT and Planck energy scales (a feature, for example, of string theories and various solutions to the hierarchy problem).

Improving the sensitivity of spin-based sensors will extend the reach of such experiments to higher energy scales as coupling constants typically scale proportionally to $1/M$ or $1/f_a$.
Spin-based sensors can also be used as particle detectors by precisely measuring and characterizing changes to the environment caused by new particle interactions. 
Because precision experiments are often carried out at the ``table-top'' scale involving relatively small teams of researchers and relatively fast timelines from conception to data, they offer affordable opportunities to explore many creative theoretical scenarios of beyond-the-Standard-Model physics.

In terms of technological development of instrumentation essential for expanding the reach of precision spin-based sensors for fundamental physics research, there are a number of high priority areas:
\begin{itemize}
    \item Find ways to enhance the number of polarized spins $N$ via optical pumping and other hyperpolarization methods and quantum control techniques, as the shot-noise-limited sensitivities of spin-based sensors generally scale proportionally to $1/\sqrt{N}$ \cite{auzinsh2004can};
    \item Develop methods and find systems to achieve the longest possible spin coherence times $\tau$, since measurement sensitivity generally scales as $1/\sqrt{\tau}$ \cite{auzinsh2004can};
    \item Improve fundamental sensitivity of spin-based sensors via new measurement schemes involving, for example, quantum back-action evasion and rapid averaging of quantum uncertainty in highly correlated spin systems (e.g., ferromagnets);
    \item Study new atomic, molecular, and condensed-matter systems that feature enhanced sensitivity to beyond-the-Standard-Model physics, such as non-centrosymmetric crystals, polyatomic molecules, and deformed nuclei;
    \item Advance tools to control and eliminate systematic errors and spurious technical noise, such as comagnetometers and quantum sensor networks;
    \item Find techniques to increase the bandwidth of spin-based sensors to explore higher frequencies and therefore higher boson masses in dark matter haloscope searches;
    \item Develop methods to speed up the scanning rate of magnetic-resonance-based dark matter haloscope searches in order to explore larger ranges of boson masses over a given measurement time;
    \item Design and implement new strategies for spin-based sensors at smaller length scales to probe higher mass exotic bosons that mediate forces with smaller length scales;
    \item Enhance the accuracy of spectroscopic measurements and theoretical calculations of atomic, molecular, and nuclear systems to enable new tests of fundamental interactions.
\end{itemize}

\section{Science targets}
\label{sec:science-targets}

Measurements of spins can probe new physics in three primary ways: 
\begin{itemize}
    \item{First, new physics may break symmetries of the Standard Model, giving rise to novel responses of Standard Model spins to other Standard Model fields (Sec.~\ref{sec:EDMs}).}
    \item{Second, the new physics may directly affect the spin, for example, via an interaction between a new field and the spin (Secs.~\ref{sec:MagsComags} -- \ref{sec:NMR}).}
    \item{Third, the environment of the spin may be affected by the new physics and the spin can discover the new physics by sensing changes to its environment (Sec.~\ref{sec:defects}).}
\end{itemize}

The canonical science target for the first kind of effect, namely, the breaking of Standard Model symmetries by new physics, is the search for the electric dipole moment of fundamental particles. If fundamental particles carried an electric dipole moment, an applied electric field will cause the spin of fundamental particles to precess. Such a dipole moment violates CP symmetry (the combined symmetry of charge conjugation, C, and parity, P) and it is a natural facet of many theories of physics beyond the Standard Model. Indeed, the existence of such CP violation is indicated by the existence of the matter-antimatter asymmetry in the universe. 

Key science targets that cause the second kind of effect, namely, direct effects on the spin itself, include particles such as axions, axionlike particles (ALPs), massive vector bosons and other ultra-light bosons. Particles of this kind emerge in several theoretical frameworks that are aimed at solving outstanding problems of the Standard Model such as the strong CP and hierarchy problems. They are also predicted to emerge as a generic consequence of string theory. The key reason for the ubiquity of such particles in these extensions of the Standard Model is due to  effective field theory. Given a light field, interactions with the spin of Standard Model fermions is one of the dominant operators that would allow this light field to interact with the Standard Model in a technically natural way. These fields are thus natural portals into the ultra-violet.  They can be detected by sourcing them in the laboratory with test masses or by looking for a cosmological abundance of them. The latter possibility is  well motivated since many cosmological scenarios (such as inflation) naturally produce a cosmic abundance of these particles. If discovered, these particles thus have the potential of solving both the problem of dark matter as well as unveiling the mysteries of the early universe. In addition, it is also possible that complex dark sectors could directly source these long range fields giving rise to new long range interactions between the dark matter and Standard Model spins. In light of poor observational constraints on the masses of such particles, it is vital to invest in technological probes that are able to cover wide swaths of parameter space. The extraordinary developments in QIS technologies over the past decade has now made such a broad probe of parameter space experimentally feasible. 

Science targets for the third possibility, namely, the use of spins to detect the effects of new physics on the environment of the spin, includes the detection of crystal damage caused by dark matter interactions and the ability to use spins to detect changes caused to surfaces at the single atom level, with the changes being produced as a result of dark matter interactions. The former phenomenon could conceivably be used to identify the direction of dark matter induced nuclear recoil while the latter could potentially be used to detect light dark matter. 

\section{Searches for parity- and time-reversal-violating \\ electric dipole moments (EDMs)}
\label{sec:EDMs}

The first way that precision measurements of spin dynamics can probe new physics identified in Sec.~\ref{sec:science-targets} is via searches for discrete symmetry violations. The primary focus of recent research has been measurement of \emph{permanent} electric dipole moments (EDMs) in atomic, molecular, and nuclear systems.
There have been a number of reviews on the topic of EDMs, see for example Refs.~\cite{demille2017probing,safronova2018search,khriplovich2012cp,ginges2004violations,pospelov2005electric,commins2007electric,commins2007EDM,chupp2010permanent,engel2013electric,jungmann2013searching,chupp2019electric}.
A nonzero EDM $\bs{d}$ of an elementary or composite particle must be proportional to the total angular momentum $\bs{F}$ of the system (a fact that follows from the Wigner-Eckart theorem and the fact that no additional quantum numbers are required to describe the system, see, for example, Refs.~\cite{khriplovich2012cp,budker2008atomic}). 
Since $\bs{d}$ is odd with respect to mirror-symmetry (parity, P) and even under time-reversal (T) while $\bs{F}$ is even under P and odd under T, the existence of an EDM violates P and T symmetries. 
Thus an EDM is a result of P- and T-violating fundamental interactions, and assuming CPT invariance, CP-violating interactions. 
Such symmetry-violating interactions can endow elementary particles such as electrons and quarks with EDMs, which can in turn create EDMs of atoms, molecules, and nuclei. Symmetry-violating interactions between constituent particles of composite systems can also induce electrical polarization along $\bs{F}$ and generate EDMs. 

The predominance of matter over antimatter is incompatible with Standard Model mechanisms of baryogenesis \cite{shaposhnikov1987baryon}, and it is widely believed that the missing ingredient is a new, larger source of CP violation that would also generate EDMs. 
A wide variety of beyond-the-Standard-Model theories predict EDMs near present experimental sensitivities. 
For instance, existing experimental limits have established some of the most stringent constraints on supersymmetric theories, in many scenarios beyond constraints from collider experiments \cite{chupp2015electric}. 
A companion Snowmass white paper, Ref.~\cite{SnowmassEDM}, provides a comprehensive review of EDM experiments. 
In the present work we provide a summary of the key experimental technology as it relates to precision measurements of spin dynamics.

Depending on whether the atomic or molecular system studied is \emph{paramagnetic} (with unpaired electron spins) or \emph{diamagnetic} (with closed electron shells but nonzero nuclear spin), different types of physics can be probed: EDM experiments with paramagnetic systems can target electron EDMs and CP-violating electron-nucleon interactions; diamagnetic systems can target nuclear EDMs and CP-violating hadronic and other semileptonic interactions.
Thus it is valuable to develop techniques and experiments to study both paramagnetic and diamagnetic systems.

The general approach of EDM experiments is to search for the combined effect of a P- and T-odd Hamiltonian and an applied electric field $\bs{E}$, which results in an energy shift $\pm\Delta E\ts{edm}$ for a given quantum state of the atom or molecule, where the sign of the effect depends on the projection of the spin along the quantization axis.
A preliminary consideration is that in the nonrelativistic limit there is no energy shift when $\bs{E}$ is applied to a neutral system composed of particles possessing nonzero EDMs. 
This is because particles will rearrange upon application of the applied field $\bs{E}$ so that the system's internal field $\bs{E}\ts{int}$ cancels $\bs{E}$ at the positions of the constituent particles, a result known as Schiff's theorem \cite{schiff1963measurability}.
However, relativistic effects not only evade Schiff's theorem but can even lead to enhancement of EDM observables \cite{commins2007electric,sandars1965electric}.
Because relativistic effects are more prominent in heavy atoms, $\Delta E\ts{edm}$ can be significantly enhanced in systems with large atomic number $Z$ \cite{commins2007electric,sandars1965electric}, and thus EDM experiments employ heavy atoms such as Tl, Th, Cs, Hg, and Xe.
Typically the system is spin polarized via optical pumping or some other hyperpolarization technique such that the system is an a superposition of quantum states with opposite EDM-induced energy shifts.
Thus a nonzero EDM will cause the polarized spins to precess in the presence of $\bs{E}$ by an angle $\phi = 2 \Delta E\ts{edm} \tau / \hbar$, where for maximum precession the time $\tau$ is given by the spin-coherence time.
The minimum achievable energy resolution for a single-particle measurement is $\hbar / (4 \tau)$; measuring with $N$ uncorrelated systems for a total time $T$ gives an energy resolution of $\delta E$
\begin{align}
\delta E \approx \frac{\hbar}{4}\frac{1}{\sqrt{ \tau T N }}~.
\label{eq:EDM-sensitivity}
\end{align}

Considering this general approach, there are several general areas of technological development that can advance the fundamental sensitivity of EDM measurements: 
\begin{itemize}
    \item{increase $\Delta E\ts{edm}$ by finding atomic and molecular systems with maximal enhancement factors;} 
    \item{improving hyperpolarization and quantum control techniques so that the total number $N$ of polarized atoms/molecules can be increased;}
    \item{achieve longer spin-coherence times $\tau$.}
\end{itemize}
At least equally important is improving control of systematic errors that could mimic EDM signals. 
Among the most pernicious systematic effects that have plagued generations of EDM experiments are those due to uncontrolled magnetic fields $\bs{B}$ that couple to the magnetic dipole moments of the atoms or molecules, causing Larmor precession of spins.
While many magnetic field effects can be distinguished from effects due to EDMs by reversal of the direction of $\bs{E}$, there can be $\bs{B}$-fields correlated with the direction of $\bs{E}$ due to leakage currents as well as motional magnetic fields $\propto \bs{E} \times \bs{v}/c$, where $\bs{v}$ is the particle velocity in the lab frame.
Magnetic-field-related systematic errors are generally reduced using the technique of \emph{comagnetometry} \cite{terrano2021comagnetometer}, where simultaneous measurements in the same volume are carried out on either different species \cite{regan2002new} or different quantum states of the same species \cite{eckel2013search}.

In addition to comagnetometry, controlling and monitoring the magnetic environment of an EDM experiment should also make use of ultra-stable and sensitive magnetometers surrounding the experiment. 
Optically pumped atomic magnetometers \cite{budker2007optical,bud13OpticalMagnetometry} and in general magnetometers probed by light are subject to shifts that challenge their stability. Nuclear spin magnetometers, in particular those based on $^3$He \cite{farooq2020absolute,abel2019picotesla}, have the potential as quantum sensors to provide the unprecedented stability required for future EDM experiments.

Earlier generations of electron-EDM experiments generally employed paramagnetic atomic systems like Cs \cite{murthy1989new} and Tl \cite{regan2002new,commins1994improved}, and there are ongoing atomic EDM experiments employing advances in laser-cooled and trapped atoms and other state-of-the-art QIS methods \cite{zhu2013absolute,wundt2012quantum,inoue2015experimental}. 
However, in recent years the focus has shifted to molecular systems such as YbF \cite{hudson2011improved}, ThO \cite{acme2014order,andreev2018improved}, and HfF$^+$ \cite{cairncross2017precision}. 
The molecular systems have enabled orders-of-magnitude improvements in sensitivities to electron EDMs through their larger enhancement factors which increase $\Delta E\ts{edm}$ as compared to atomic systems as well as opening a variety of techniques to control and reduce systematic errors. 
Efficient systematic error control in molecular EDM experiments is accomplished by experimenting on particular molecular states that have reduced sensitivities to magnetic perturbations while retaining sensitivity to EDM-induced effects, and by using optical and radio-frequency fields for quantum control to switch between different quantum states that allow rapid measurement and cancellation of many systematic errors.
Further improvements in cooling \cite{carr2009cold} and control of molecules \cite{koch2019quantum,bohn2017cold}, extending spin-coherence times \cite{caldwell2020long,zhou2020second}, increasing the number of polarized molecules \cite{julienne2018quo,wu2020metastable,anderegg2017radio}, and advances in comagnetometry \cite{kozyryev2017precision} and other methods to control systematic effects are among the paths toward further advances.
In addition to ongoing experiments \cite{hudson2011improved,acme2014order,andreev2018improved,cairncross2017precision}, a number of new experiments are under development \cite{aggarwal2018measuring}.

The leading diamagnetic (nuclear) EDM experiment has employed Hg atoms \cite{graner2016reduced}, complementary to direct measurements of the neutron EDM \cite{abel2020measurement}.
The sensitivity of the Hg EDM experiment results from a relatively high density of optically polarized atoms ($N \sim 10^{14}$) and long coherence times (hundreds of seconds), as well as a variety of auxiliary measurements and techniques developed over the years to reduce systematic errors \cite{swallows2013techniques}. 
Searches for EDMs of diamagnetic atoms in other systems have been carried out \cite{cho1991search,parker2015first,bishof2016improved,sachdeva2019new,allmendinger2019measurement}; many of these are ongoing efforts with the prospect of improving measurement accuracy by orders of magnitude, such as the radium EDM search in which several upgrades are in the process of being implemented \cite{booth2020spectroscopic,ready2021surface}.
There are also a number of new experiments that have the potential to explore unconstrained parameter space for symmetry violating effects in the nuclear sector \cite{kozyryev2017precision,singh2019new,maison2019theoretical,denis2020enhanced,jadbabaie2020enhanced,garcia2020spectroscopy,fan2021optical,yu2021probing}, such as the CENTREX experiment that employs a cold beam of TlF molecules \cite{grasdijk2021centrex}, a search particularly sensitive to the proton EDM \cite{sandars1967measurability,wilkening1984search}.

Technological improvements that can enhance the sensitivity of EDM experiments include any methods that result in longer spin-coherence times, such as longer beam lines, slower/colder beams, and trapping of molecules which can lengthen spin-coherence times by orders of magnitude. 
Sensitivity can also be improved by increasing count rates via beam cooling and focusing, more efficient probing/detection methods, improved trapping techniques, and brighter molecular sources. 
It is important to note that all three of the leading electron EDM searches with molecules \cite{hudson2011improved,andreev2018improved,cairncross2017precision} are presently statistics limited, meaning that technological advances in the aforementioned areas can lead directly to improved sensitivity.

An important area of technological development is toward the use of deformed nuclei for EDM searches \cite{auerbach1996collective}.
Because the motion of a nucleus within an atom or molecule is deeply nonrelativistic, Schiff's theorem \cite{schiff1963measurability} implies that any nuclear EDM is mostly screened from external fields.
Nonetheless, symmetry violating nuclear interactions can change the nuclear charge and current distributions, and lead to nonzero energy shifts described by the Schiff moment \cite{flambaum2002nuclear}.
Deformed nuclei that possess a reflection antisymmetric shape, such as Fr, Ra, Th, and Pa that have static octopole deformations, have enhanced nuclear Schiff moments (by orders of magnitude) and therefore lead to comparably larger atomic and molecular EDMs \cite{auerbach1996collective,spevak1997enhanced,chishti2020direct}.

Another rapidly developing technology, useful not only for nuclear EDM experiments but also for a wide range of searches for beyond-the-Standard-Model physics, are new methods for nuclear spin comagnetometry \cite{sachdeva2019new,ledbetter2012liquid,wu2018nuclear,limes2018he}. These techniques can improve control of systematic errors, often the limiting factor in EDM experiments.

A new direction of particular interest is the use of polyatomic molecules for EDM searches, which can enable application of laser cooling techniques \cite{mitra2020direct} in conjunction with internal comagetometry and full polarization \cite{kozyryev2017precision,hutzler2020polyatomic}. Polyatomic molecules show considerable promise for both electron and nuclear EDM experiments.

A different approach is to develop solid-state systems for EDM experiments \cite{budker2006sensitivity,leggett1978macroscopic}.
Such solid-state EDM experiments sacrifice the long spin-coherence times possible in gas-phase atomic and molecular experiments for a significantly larger signal due to the higher density of spins in a solid.
As first suggested in Refs.~\cite{Sha68EDMs,ignatovich1969amplification}, an electron EDM search can be carried out using unpaired election spins bound to a crystal lattice: when an electric field $\bs{E}$ is applied, if the electrons possess a non-zero EDM the spins will become oriented parallel to $\bs{E}$ and generate a nonzero magnetization \cite{vasil1978measurement,kim2011experimental,eckel2012limit}.
The inverse experiment can also be performed, where a material is magnetized (spin-polarized) and one searches for electric polarization due to a nonzero electron EDM \cite{heidenreich2005limit}.
Technological improvements are needed to reduce systematic errors in such solid-state EDM experiments, for example due to heating and dielectric relaxation.

In the longer term, it is likely that advances along multiple fronts will allow the frontiers of EDM searches to be pushed even further. For example, using heavy polar molecules with deformed nuclei in an EDM experiment taking full advantage of state-of-the-art cooling, trapping, and molecular production could allow sensitivity to symmetry-violating interactions many orders of magnitude beyond what is possible today \cite{yu2021probing}. 
Combining the Schiff moment enhancement of an octopole-deformed nucleus with the relativistic enhancement, there are molecular species such as $^{229}$ThO, $^{229}$ThOH, $^{229}$ThF$^+$, and $^{225,223}$RaOH$^+$, $^{225,223}$RaOCH$^+_3$, $^{225,223}$RaF, $^{225,223}$RaAg, and $^{223}$FrAg that are up to $10^6$ times more sensitive per particle to CP-violating physics than $^{199}$Hg \cite{sushkov1984possibility,flambaum2019enhanced,klos2022prospects}.
Note that dedicated institutes for low-energy nuclear science research, such as the Facility for Rare Isotope Beams (FRIB), TRI-University Meson Facility (TRIUMF), and Isotope mass Separator On-Line (ISOLDE), have the capability to produce these isotopes for use in practical quantities and enable precursor spectrocopic studies \cite{abel2019isotope}.
Excellent candidates are Ra-containing molecules \cite{fleig2021theoretical}, since Ra has a well-studied nuclear deformation \cite{gaffney2013studies,butler2020evolution}, and many Ra-containing molecules can be laser cooled.
For example, RaOCH$^+_3$ was recently synthesized, trapped and cooled in an ion trap \cite{fan2021optical}, opening the potential for an experiment that takes advantage of the advanced quantum control techniques possible with cold ions \cite{cairncross2017precision,yu2021probing,chou2017preparation}.
A novel related concept is to use the radioactive species $^{229}$Pa, which may be a highly deformed nucleus, embedded in an optical crystal to search for its strongly enhanced symmetry-violating magnetic quadrupole moment \cite{singh2019new}.

Another route is to combine the advantages of the long coherence times and quantum control possible in gas-phase atomic and molecular experiments with the high spin densities possible in solid-state systems \cite{pryor1987artificial,arndt1993can,kozlov2006proposal,xu2011optical,upadhyay2016longitudinal,vutha2018oriented,vutha2018orientation}.  
The idea is to trap atoms and molecules with high intrinsic sensitivity to symmetry-violating interactions within inert cryogenic crystal matrices.
In order for an EDM experiment based on this approach to surpass the sensitivities of gas-phase experiments, it is essential that both high density of the target species is achieved and that the target species retains all the key properties that enable quantum control and sensing in the inert crystal environment (long coherence times and efficient polarization and read-out of spin states).
While experiments with alkali atoms in solid hydrogen and solid helium have demonstrated long coherence times and efficient optical pumping and probing \cite{arndt1995long,kanorsky1996millihertz,lang1999optical}, the alkali atom densities so far have been low.
On the other hand, both high alkali atom density and relatively long spin-coherence times ($\tau \equiv T_2 \sim 0.1$~s) have been demonstrated in solid parahydrogen \cite{upadhyay2019enhanced,upadhyay2019spin,upadhyay2020ultralong}.
While there are experimental hurdles yet to be overcome, such as relatively short spin ensemble dephasing times ($T_2^*$) due to the polycrystalline
nature of the parahydrogen samples used so far \cite{upadhyay2019enhanced,upadhyay2019spin}, there are viable paths forward to taking full advantage of the possibilities of this system by, for example, creating single-crystal cryogenic samples \cite{batchelder1967measurements}.

Many of the EDM experiments described here rely on quantum sensing and control of spin ensembles, analogous to those used in QIS, and can therefore borrow new tools from this rapidly-advancing field \cite{demille2002quantum,koch2019quantum,henriet2020quantum}. 
It is widely believed that new sources of CP-violation are required to explain the cosmological matter-antimatter asymmetry \cite{dine2003origin}, and, consequently, the wide range of beyond-the-Standard-Model theories predicting observable EDMs ``just around the corner'' of present experimental sensitivities \cite{chupp2015electric}.
Discovery of a nonzero EDM would herald the existence of new particles, and can explore new physics from particles with masses beyond the direct reach of any conceived accelerator \cite{demille2017probing,safronova2018search}.
Therefore this area of research is of highest priority for the precision measurement community.

\section{Searches for exotic spin-dependent interactions \\ using magnetometry and comagnetometry}
\label{sec:MagsComags}

The second class of precision experiments highlighted in Sec.~\ref{sec:science-targets} are direct searches for exotic spin-dependent interactions originating from beyond-the-Standard-Model physics. 
Many theories predict the existence of new force-mediating bosons that couple to the spins of Standard Model particles \cite{safronova2018search}.
Regardless of the specifics of the fundamental theory, if the new interaction respects rotational invariance, there are only a relatively small number of long-range interaction potentials that can exist as described in detail in Refs.~\cite{moody1984new,dobrescu2006spin,fadeev2019revisiting}.
The range of such a fundamental interaction is parameterized by the Compton wavelength of the force-mediating boson: $\lambda_c = \hbar/\prn{ mc }$, where $m$ is the boson mass.
For example, exchange of an exotic spin-0 boson (such as an axion \cite{moody1984new}) with pseudoscalar coupling to fermion $1$ and scalar coupling to fermion $2$ leads to a monopole-dipole potential of the form:
\begin{align}
\mc{V}_{ps}(r) = \frac{g_p^{(1)} g_s^{(2)}  \hbar}{8 \pi m_1 c} \bs{S}_1 \cdot \hat{\bs{r}} \prn{ \frac{1}{r\lambda_c} + \frac{1}{r^2} }  e^{-r/\lambda_c}~,
\label{Eq:Vps}
\end{align}
where $g_p^{(1)}$ and $g_s^{(2)}$ parameterize the vertex-level pseudoscalar and scalar couplings, respectively, $\bs{S}_1$ is the spin of fermion $1$, $m_1$ is mass of fermion $1$, and $\bs{r} = r\hat{\bs{r}}$ is the displacement vector between the fermions.
The potential $\mc{V}_{ps}(r)$ causes an associated spin-dependent energy shift.
The basic experimental program is thus to hunt for all possible types of interactions at various length scales between Standard Model fermions (typically electrons, protons, and neutrons in the case of AMO experiments).
Through the framework of Refs.~\cite{moody1984new,dobrescu2006spin,fadeev2019revisiting}, the results of experiments can be interpreted in terms of fundamental physics theories \cite{safronova2018search}.

One of the primary experimental strategies is to employ a sensitive detector of torques on spins and then bring that spin-based torque sensor within $\sim \lambda_c$ of an object that acts as a local source of an exotic field (e.g., a large mass or highly polarized spin sample).
Such experiments are closely analogous to spin-based magnetometry \cite{budker2007optical,bud13OpticalMagnetometry}, where the effect of an ambient magnetic field $\bs{B}$ is measured by sensing the $\bs{\mu}\times\bs{B}$ torque on spins with magnetic moment $\bs{\mu}$. 
This is equivalent to measuring the magnetic-field-induced energy shift between Zeeman sublevels via observation of the time-evolution of a coherent superposition of spin states in the probed system.
Exotic spin-dependent interactions act as ``pseudo-magnetic fields'' and generate analogous effects, albeit with couplings to Standard Model particles that can be completely different from those from a real magnetic field \cite{kimball2015nuclear,kimball2016magnetic}.

\begin{figure}
\center
\includegraphics[width=14cm]{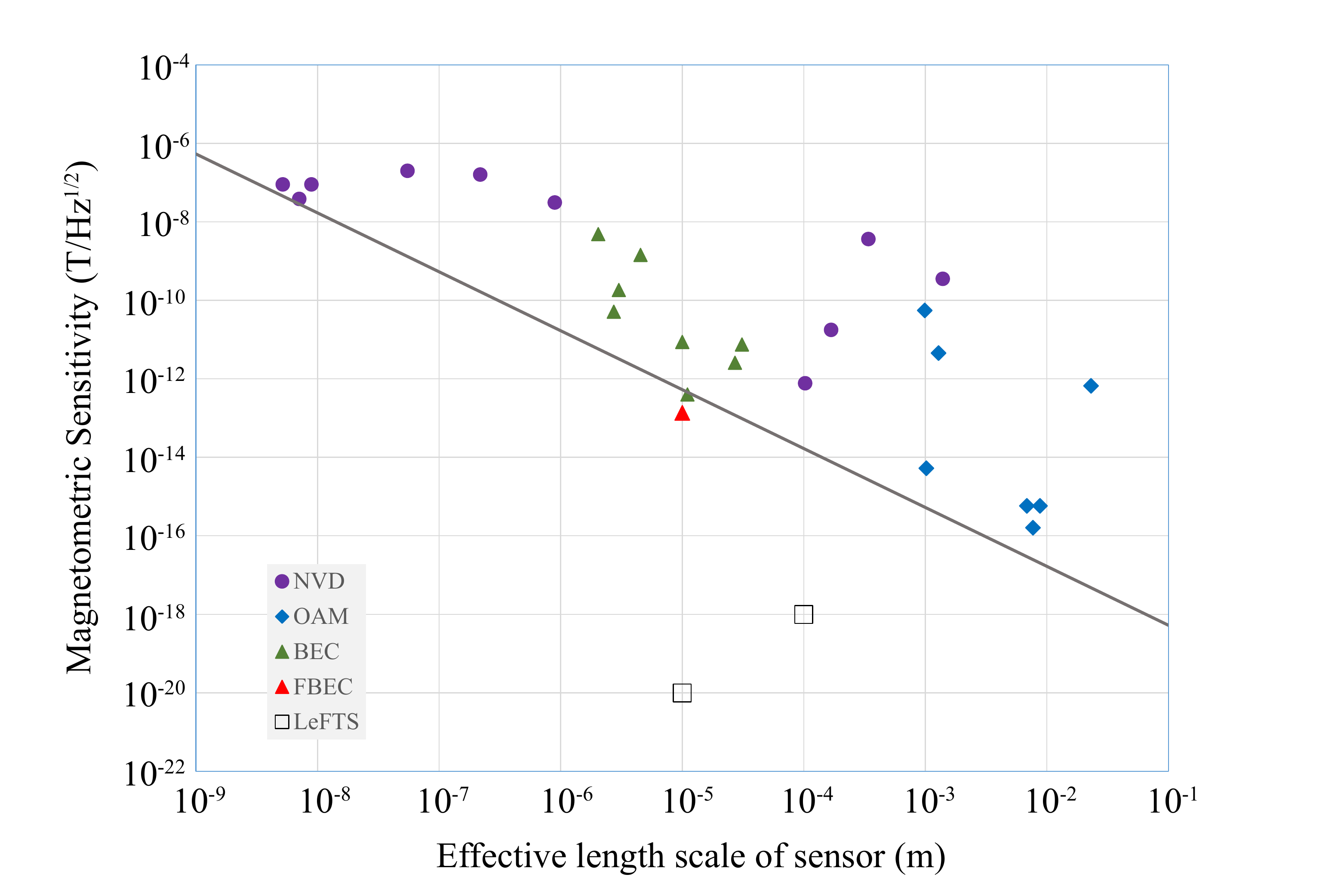}
\caption{Summary of the size and sensitivity of spin-based magnetometers. Experimentally demonstrated magnetometers are represented by filled markers, projected sensitivity of proposed magnetometers are represented by unfilled markers. The gray line indicates the energy resolution limit (ERL) described by Eq.~\eqref{eq:ERL}. The purple circles correspond to nitrogen-vacancy diamond (NVD) magnetometers \cite{fang2013high,trusheim2014scalable,clevenson2015broadband,lovchinsky2016nuclear,barry2016optical,ahmadi2017pump,zhou2020quantum}, the green triangles correspond to atomic Bose-Einstein condensate (BEC) magnetometers \cite{wildermuth2005microscopic,wildermuth2006sensing,vengalattore2007high,ockeloen2013quantum,eto2013spin,muessel2014scalable,wood2015magnetic,yang2017scanning,jasperse2017continuous}, and the blue diamonds correspond to optical atomic magnetometers (OAM) \cite{kominis2003subfemtotesla,schwindt2004chip,schwindt2007chip,griffith2010femtotesla,dang2010ultrahigh,sewell2012magnetic,sheng2013subfemtotesla,behbood2013real}. The red triangle represents the sensitivity of the recently demonstrated single-domain ferromagnetic BEC magnetometer (FBEC) that surpasses the ERL \cite{palacios2022single}. Levitated ferromagnetic torque sensors (LeFTS), represented by the unfilled black squares, are predicted to surpass the ERL by many orders of magnitude \cite{kimball2016precessing,vinante2021surpassing}. Figure adapted from Ref.~\cite{mitchell2020colloquium}; does not include non-spin-based magnetic sensors based on, for example, superconducting quantum interference devices (SQUIDs).}
\label{Fig:sensitivity-magnetometers}
\end{figure}

The central technology in these experiments is the spin-based sensor employed.
The parameter space accessible depends on the overall sensitivity, which determines how small a coupling can be observed, as well as the size and geometry of the sensor, which determines what interaction range $\lambda_c$ (boson mass $m$) can be probed.
Since the observable in these experiments is a spin-dependent energy shift, just as in the case of the EDM experiments discussed in Sec.~\ref{sec:EDMs}, a sensor employing $N$ independent spins with coherence time $\tau$ has a shot-noise-limited sensitivity described by Eq.~\eqref{eq:EDM-sensitivity}.
However, as noted in Ref.~\cite{mitchell2020colloquium}, a practical benchmark for comparison of different magnetometer technologies is the {\emph{energy resolution limit}} (ERL) given by
\begin{align}
\frac{\prn{\delta B}^2 V T}{2\mu_0} \gtrsim \hbar~,
\label{eq:ERL}
\end{align}
where $\delta B$ is the magnetometric sensitivity per root Hz, $V$ is the sensing volume, $T$ is the measurement time, and $\mu_0$ is the vacuum permeability.
Heuristic arguments for the ERL are given in Refs.~\cite{mitchell2020colloquium}: as can be seen from Eq.~\eqref{eq:ERL}, the ERL is based on equating the energy associated magnetic field fluctuations in volume $V$ over time $T$ with $\hbar$.
The ERL, while not a fundamental limit, is empirically a difficult level for magnetometers to surpass, as can be seen from the survey of current technology shown in Fig.~\ref{Fig:sensitivity-magnetometers}.

Therefore, a major technological leap in the search for exotic spin-dependent interactions at various length scales would be to find methods to surpass the ERL. 
One promising technology along these lines is the development of levitated ferromagnetic torque sensors (LeFTS) \cite{kimball2016precessing,vinante2021surpassing,wang2019dynamics,gieseler2020single,vinante2020ultralow,fadeev2021gravity,fadeev2021ferromagnetic}.
The active sensing element consists of a hard ferromagnet, well isolated from the environment by, for example, levitation over a superconductor via the Meissner effect.
The mechanical response of the levitated ferromagnet to an exotic spin-dependent interaction can be precisely measured using a superconducting quantum interference device (SQUID).
For sufficiently slow rotational motion, the ferromagnet's angular momentum is dominated by its intrinsic spin, and it behaves as a gyroscope \cite{kimball2016precessing}.
For faster motion, the levitated ferromagnet's dynamics are dominated by pendulum-like librational motion \cite{vinante2021surpassing}.
In either regime, LeFTS are predicted to be able to surpass both the ERL and even the standard quantum limit (SQL) for uncorrelated spins described by Eq.~\eqref{eq:EDM-sensitivity}.
The ability of LeFTS to achieve this sensitivity is a result of the high correlation of the electron spins in a ferromagnet, which are locked together along a well-defined local direction by magnetic anisotropy, ultimately converting the field measurement into a mechanical measurement \cite{vinante2021surpassing}.
The quantum uncertainty in the spin orientation is rapidly averaged by the strong internal interactions in the ferromagnet \cite{kimball2016precessing}.

Recently, a magnetic field sensor surpassing the ERL was demonstrated: a single domain spinor Bose-Einstein condensate (BEC) \cite{palacios2022single}. 
Similar to the LeFTS concept, ultracold two-body interactions in the BEC create a fully coherent, single-domain state of the atomic spins that enables the system to evade the ERL that limits traditional spin-based sensors.
The experiment described in Ref.~\cite{palacios2022single} confirms the principles underlying the promise of next-generation torque sensors such as LeFTS.

A variety of other directions to improve fundamental and practical sensitivity of spin-based magnetometers are being explored, including bandwidth enhancement via spin squeezing \cite{shah2010high,wasilewski2010quantum,pezze2018quantum} and methods to utilize many-body collective correlation among spins \cite{shuker2021photon}.
A new high-frequency magnetometer based on electron spin resonance, operating in the MHz~--~GHz region, has demonstrated precision at the pT level and has the potential to reach sub-fT sensitivity \cite{crescini2021phase}. 

Beyond the intrinsic sensitivity, the principal challenge in experiments searching for exotic spin-dependent interactions is understanding and eliminating systematic errors: clearly distinguishing exotic spin-dependent interactions from mundane effects due to, for example, magnetic interactions.
This is a theme in common with the EDM searches discussed in Sec.~\ref{sec:EDMs}, and many similar technical approaches avail themselves.
Ideally, the local source of the exotic field can be manipulated in such a way as to modulate its effects, thereby providing a signal with a well-characterized time-dependence that can be distinguished from background.
In addition, a variety of independent measurements can be used to monitor, control, and identify systematic errors.
Importantly, in searches for exotic spin-dependent potentials, the sought-after effect is not due to a real magnetic field, but rather a {\emph{pseudo-magnetic field}}.
Therefore, by comparing the response of two different systems, effects from magnetic fields can be distinguished from effects due to exotic spin-dependent interactions.
This is the essence of {\emph{comagnetometry}} \cite{lamoreaux1989electric}, where the same field, magnetic or otherwise, is simultaneously measured using two different ensembles of atomic or nuclear spins, reviewed in Ref.~\cite{terrano2021comagnetometer}.

Comagnetometers are in fact the most sensitive devices for measuring energy differences between quantum states, in some cases achieving precision at the $\sim 10^{-26}~{\rm eV}$ level \cite{vasilakis2009limits,brown2010new,allmendinger2014new}.
Presently the most sensitive alkali-atom/noble-gas comagnetometers are based on {\emph{spin-exchange-relaxation-free}} (SERF) atomic magnetometry combined with a scheme where the magnetization of a noble gas species self-compensates the magnetic field, and enabling nearly background-free searches for exotic spin-dependent interactions \cite{kornack2002dynamics,kornack2005nuclear}.
Other methods have reached similar sensitivity using a variety of atomic systems via simultaneous measurement of spin-precession in different samples \cite{allmendinger2014new,graner2016reduced}.

Presently, comagnetometer technology is limited by effects due to the combination of magnetic field gradients and imperfect sample overlap, atomic collisions, surface interactions that differentially affect the atomic species, and quantum back-action. 
A number of techniques to circumvent these limitations are being explored. 
For example, in Ref.~\cite{limes2018he}, quantum control methods are used to average away deleterious effects and precession is measured ``in the dark'' without external fields applied in order to reduce background effects.
In Ref.~\cite{wu2018nuclear}, the problem of magnetic field gradients is overcome in a liquid ensemble of identical molecules by carrying out comagnetometry with different nuclear spins in each identical molecule, suppressing effects of gradients by over an order-of-magnitude as compared to overlapping samples of different atoms/molecules. 

\begin{figure}
\center
\includegraphics[width=14cm]{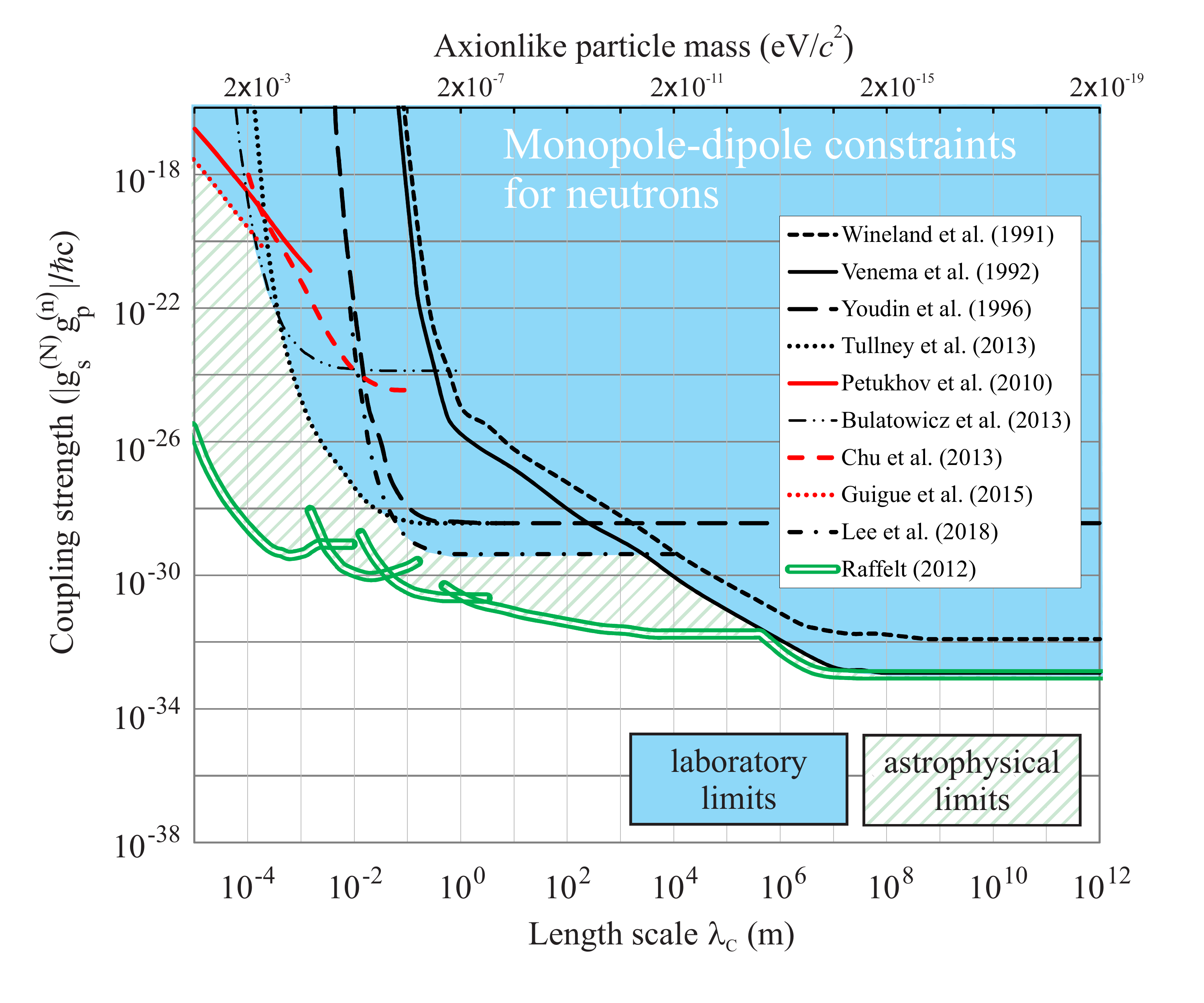}
\caption{Constraints on the axionlike-particle-mediated monopole-dipole interaction between nucleons and neutrons, $|g_p^{(n)} g_s^{(N)}|/(\hbar c)$, as described by Eq.~\eqref{Eq:Vps}, adapted and updated from Ref.~\cite{safronova2018search}. Experiments using comagnetometry \cite{Win91,Ven92,You96,Tul13,Bul13,lee2018improved} are indicated by black lines, experiments using magnetometry are indicated by red lines \cite{Pet10,Chu13,Gui15}, astrophysical constraints are indicated by the green double line \cite{Raf12}. Experiments at different length scales measure interaction ranges corresponding to different axionlike particle (ALP) Compton wavelengths $\lambda_c$, and thus different ALP masses $m$.}
\label{Fig:monopole-dipole-neutron-constraints}
\end{figure}

Magnetometer and comagnetometer technology has been applied to a wide variety of experiments searching for new spin-dependent interactions.
Experiments using spin-based sensors and spectroscopy have been able to search for interactions with ranges from the nanometer-scale \cite{Ram79,Kar10a,Kar10b,Kar10c,Led13,Fic17} to the Earth-scale \cite{Win91,Ven92,Hec08,Hun13,hunter2014using,kimball2017constraints}, and have probed interactions of protons \cite{Ram79,Led13,kimball2017constraints,salumbides2013bounds}, neutrons \cite{vasilakis2009limits,Win91,Ven92,Gle08,su2021search}, electrons \cite{ritter1993search,Ni94,hoedl2011improved,Kot15,Ter15,crescini2017improved,luo2017constraints,kim2018experimental,rong2018searching,rong2018constraints,ding2020constraints,almasi2020new,ren2021search,crescini2022search}, and even antimatter \cite{Les14,ficek2018constraints}.
To get an overall idea of the state-of-the-art in experimental methods, a representative survey of the use of spin-based sensor technology in searches for the monopole-dipole interaction described by Eq.~\eqref{Eq:Vps} for neutron spins \cite{Win91,Ven92,You96,Tul13,Bul13,lee2018improved,Pet10,Chu13,Gui15} is shown in Fig.~\ref{Fig:monopole-dipole-neutron-constraints}.
In Fig.~\ref{Fig:monopole-dipole-neutron-constraints}, constraints on the dimensionless coupling constant $|g_p^{(n)} g_s^{(N)}|/(\hbar c)$ using comagnetometers are indicated by the black lines, and constraints obtained using $^3$He magnetometry are indicated by red lines.
Parameter space excluded by laboratory experiments is indicated by the light blue shaded region and astrophysical constraints \cite{Raf12} are shown by the double green lines and green hatched region.
It is evident that the best laboratory constraints are obtained using comagnetometry techniques, and these techniques are now at the level of precision where for many boson masses they can begin to explore parameter space outside of astrophysical constraints, highlighting the importance of further technological improvements.

If $\lambda_c$ is at or below the atomic or molecular scale, experimental searches often rely on comparing high-precision measurements to high-accuracy atomic and molecular calculations based on Standard Model physics, as described, for example, in Refs.~\cite{Kar10a,Kar10b,Kar10c,Fic17,ficek2018constraints,fadeev2022pseudovector}.
The idea in these studies is that disagreement between theory and experiment can be interpreted as a possible hint of new physics, while good agreement between theory and experiment can be interpreted as a constraint on new physics scenarios.   
In these cases, improvements in spectroscopic measurement techniques must be accompanied by similar improvements in calculations, and thus there are usually advantages to studying simpler atomic and molecular systems that can be well understood.
This is a situation similar in many respects to the long-running program of atomic parity violation measurements and calculations used to test electroweak unification \cite{safronova2018search}, which, of course, can also be used to place bounds on exotic parity-violating interactions \cite{antypas2019isotopic}.
Note also that EDM measurements (Sec.~\ref{sec:EDMs}) can be used to constrain atomic- and molecular-scale symmetry violating interactions \cite{stadnik2018improved}.

\section{Spin-based sensor networks}
\label{sec:networks}

The searches for exotic spin-dependent interactions mediated by new bosons described in Sec.~\ref{sec:MagsComags} employ a local source for the new potential and a spin-based sensor to detect the effects of that potential. 
Another possibility is that the new bosons can be abundantly generated by astrophysical processes: for example, as dark matter produced in the early universe \cite{preskill1983cosmology}, or through some cataclysmic astrophysical process such as those occurring near black holes \cite{arvanitaki2011exploring,arvanitaki2015discovering,baryakhtar2021black}.
In these scenarios, the existence of the new bosons could be directly detected through their interactions with electronic or nuclear spins as reviewed in Ref.~\cite{graham2018spin}.

If exotic ultralight bosons ($m \lesssim 1~{\rm eV}/c^2$) such as axions, axionlike particles (ALPs), or dark/hidden photons make up the majority of dark matter and have negligible self-interactions, their phenomenology is well-described by a classical field oscillating at the Compton frequency $\omega_c = mc^2/\hbar$.
However, due to topology or self-interactions, such ultralight bosonic fields can form stable, macroscopic field configurations in the form of boson stars \cite{Kol93,Bra16,eby2019global} or topological defects (e.g., domain walls, strings, or monopoles \cite{Vil85}). 
Even in the absence of topological defects or self-interactions, bosonic dark matter fields exhibit stochastic fluctuations \cite{centers2021stochastic}. 
Additionally, as noted above, it is possible that high-energy astrophysical events could produce intense bursts of exotic ultralight bosonic fields \cite{dailey2021quantum}. 
In any of these scenarios, instead of being bathed in a uniform flux, terrestrial detectors will witness transient events when ultralight bosonic fields pass through Earth \cite{Pos13}.

\begin{figure}
\center
\includegraphics[width=7cm]{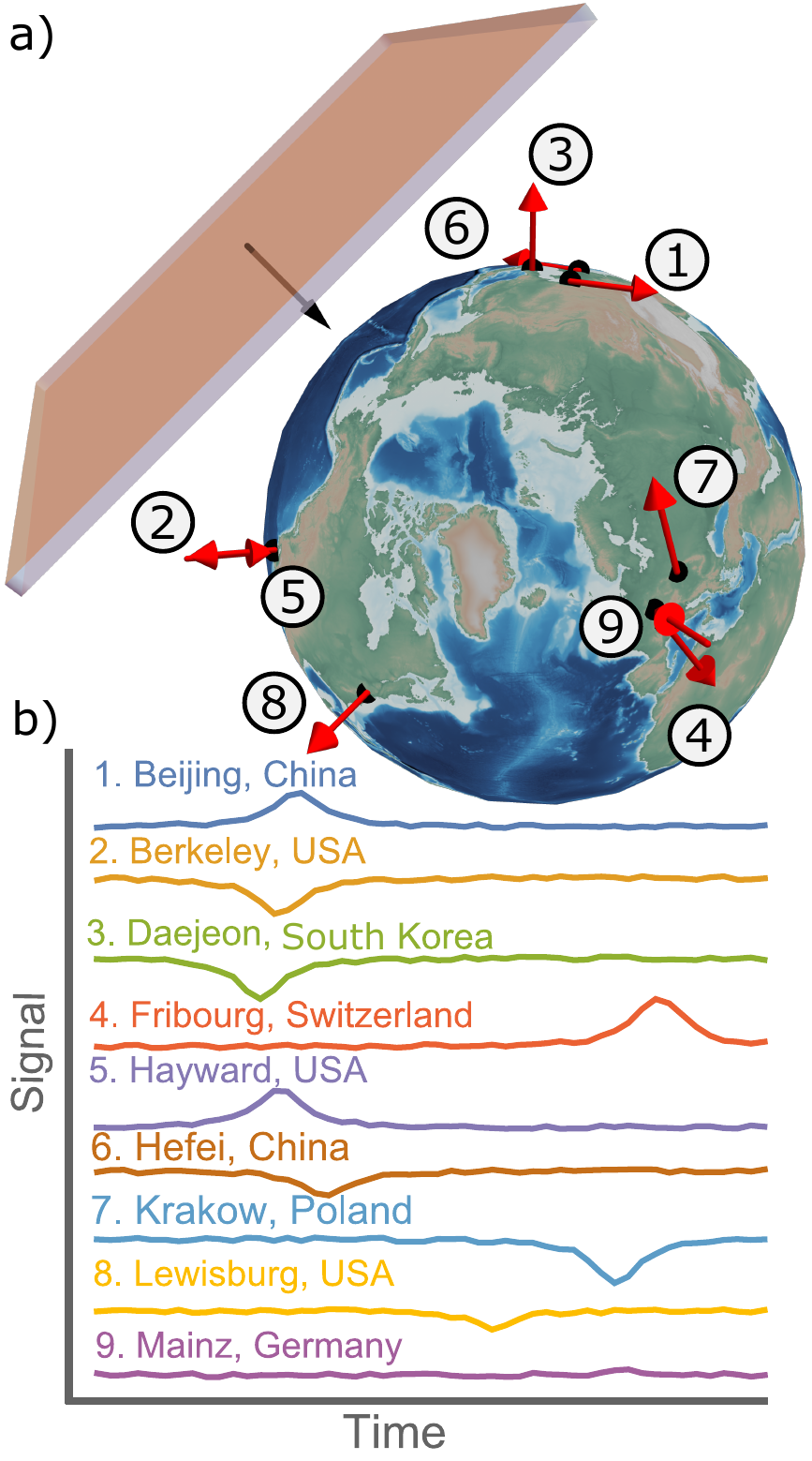}
\caption{(a) Schematic representation of a ALP field topological defect (domain wall) passing through the Earth, with the location and sensitive direction of GNOME sensors marked by arrows. (b) As the topological defect passes through various GNOME stations, signals appear in the magnetometer data at particular times. The sign and amplitude of the signals depend on the orientation of the sensor with respect to the domain wall and the atomic species used. Figure from Ref.~\cite{afach2021search}.}
\label{Fig:GNOME-domain-wall}
\end{figure}

Such transient phenomena could easily be missed by experimenters when data are averaged over long times to increase the signal-to-noise ratio as is done in the searches described in Secs.~\ref{sec:EDMs} and \ref{sec:MagsComags}.
Detecting such unconventional events presents several challenges. 
If a transient signal heralding new physics is observed with a single detector, it would be exceedingly difficult to confidently distinguish the exotic-physics signal from the many sources of noise that generally plague precision spin-based sensor measurements. 
However, if transient interactions occur over a global scale, a network of spin-based sensors geographically distributed over Earth could search for specific patterns in the timing, amplitude, phase, and polarization of such signals that would be unlikely to occur randomly, as illustrated in Fig.~\ref{Fig:GNOME-domain-wall}. 
By correlating the readouts of many sensors, local effects can be filtered away and exotic physics could be distinguished from prosaic Standard-Model physics \cite{Mas20,afach2021search,chen2021dissecting}.

This idea forms the basis for the Global Network of Optical Magnetometers for Exotic physics searches (GNOME), an international collaboration operating spin-based sensors
all over the world, specifically targeting beyond-the-Standard-Model physics \cite{Pus13,afach2018characterization}.
The magnetometric sensitivity of each GNOME sensor is $\approx~100~{\rm fT/\sqrt{Hz}}$ over a bandwidth of $\approx 100~{\rm Hz}$ \cite{afach2018characterization}. 
Each magnetometer is located within a multi-layer magnetic shield to reduce the influence of magnetic noise and perturbations while still retaining sensitivity to many exotic fields \cite{kimball2016magnetic}. 
Even with the magnetic shielding, there is inevitably some level of transient signals and noise associated with the local environment (and possibly with global effects like the solar wind, changes to the Earth's magnetic field, etc.).  
Therefore, each GNOME sensor uses auxiliary unshielded magnetometers and other sensors (such as accelerometers and gyroscopes) to measure relevant environmental conditions, enabling exclusion/vetoing of data with known systematic issues \cite{afach2018characterization}.
The signals from GNOME sensors are recorded with accurate timing provided by the global positioning system (GPS) using a custom GPS-disciplined data acquisition system \cite{Wlo14} with temporal resolution $\lesssim 10~{\rm ms}$ (determined by the magnetometer bandwidth), enabling reconstruction of events that propagate at $\lesssim c$ across the Earth ($R_E/c \approx 40~{\rm ms}$). 
The broad geographical distribution of sensors enables GNOME to achieve good spatial resolution and act as an ``exotic physics telescope'' with a baseline comparable to the diameter of the Earth \cite{dailey2021quantum}.

GNOME searches for a class of signals different from that probed by most other experiments, namely transient and stochastic effects that could arise from ALP fields of astrophysical origin passing through the Earth during a finite time. 
Depending on the particular hypothesis tested, GNOME is sensitive to ALPs with masses between $\approx 10^{-17}~\rm{eV}$ and $\approx 10^{-9}~\rm{eV}$, and can probe parameter space unconstrained by existing laboratory experiments and astrophysical observations discussed in Sec.~\ref{sec:MagsComags}.
A search for ALP domain walls has already been carried out \cite{afach2021search,kim2021machine}, and there are ongoing efforts to search for boson stars \cite{Kim18AxionStars}, carry out intensity interferometry using GNOME to detect stochastic fluctuations of dark matter fields \cite{mas2022intensity}, perform multimessenger ``exotic physics'' astronomy \cite{dailey2021quantum}, and probe other scenarios \cite{grabowska2018detecting}.
New data analysis efforts and upgrades of GNOME magnetometers to noble gas comagnetometers \cite{kornack2002dynamics,kornack2005nuclear} are underway.
Most importantly, correlated searches with spin-based sensors offer the possibility to hunt for the unexpected.

Another interesting scenario is the case of kinetically-mixed hidden-photon dark matter. 
It turns out that the Earth itself can act as a transducer to convert hidden photon dark matter into a monochromatic oscillating magnetic field at the surface of the Earth \cite{fedderke2021earth}. 
The induced magnetic field from the hidden photons has a characteristic global vectorial pattern that can be searched for with unshielded magnetometers dispersed over the surface of the Earth. 
GNOME is insensitive to such kinetically-mixed hidden-photon dark matter because of the magnetic shields enclosing the magnetometers \cite{kimball2016magnetic,Chaudhuri2015Radio}. 
Instead, a network of unshielded magnetometers is required. 
Searches for dark/hidden photons and ALPs using a publicly available dataset from the SuperMAG Collaboration \cite{fedderke2021search,arza2021earth} established experimental constraints on such scenarios that are competitive with astrophysical limits \cite{payez2015revisiting,mcdermott2020cosmological,wadekar2021gas} and the CAST experiment \cite{collaboration2017new} in the probed mass ranges (from around $10^{-18}$~eV to $10^{-16}$~eV).
A dedicated unshielded magnetometer network targeting hidden photon dark matter may be able to extend the probed parameter space.

\section{Magnetic resonance searches for ultralight bosonic dark matter fields}
\label{sec:NMR}

In contrast to the some of the scenarios discussed in Sec.~\ref{sec:networks}, the simplest assumption for the nature of ultralight ($m \lesssim 1~{\rm eV}/c^2$) bosonic dark matter postulates that the bosons are virialized in the gravitational potential of galaxies such as the Milky Way and manifest as classical fields oscillating at the Compton frequency $\omega_c$.
The bosonic dark matter field can cause spin precession via couplings to nuclear and electron spins, and since the field oscillates at a particular frequency the broad and versatile tools of {\emph{magnetic resonance}} can be used to detect the spin interaction.

An axion (or ALP) field $a(\bs{r},t)$, which to be dark matter must be nonrelativistic, can be described approximately by
\begin{align}
    a(\bs{r},t) = a_0 \cos\prn{ \bs{k}\cdot\bs{r} - \omega_c t + \phi_0}~,
    \label{eq:axion-field}
\end{align}
where $\bs{k} \approx m \bs{v} /\hbar$ is the axion wave vector ($\bs{v}$ is the relative velocity between the sensor and the field), $\phi_0$ is a random phase offset, and $a_0$ is the average field amplitude, which can be estimated by assuming the average energy of the axion field comprises the totality of the local dark matter energy density $\rho_{dm} \approx 0.4~{\rm GeV/cm^3}$
\begin{align}
    \abrk{ a_0^2 } \approx \frac{2\hbar^2}{c^2}\frac{\rho_{dm}}{m^2}~.
    \label{eq:axion-field-amplitude}
\end{align}
The axion field has a finite coherence time due to the random kinetic energy of the constituent axions, leading to a broadening of the line shape to a part in $\sim 10^6 \approx c^2/v^2$ as discussed in Refs.~\cite{krauss1985calculations,gramolin2022spectral}, as well as stochastic amplitude fluctuations \cite{centers2021stochastic}.

The canonical axion of quantum chromodynamics (QCD), a consequence of the Peccei-Quinn mechanism introduced to solve the strong-CP problem \cite{Pec77a,Pec77b}, naturally couples to the gluon field and generates an oscillating EDM $\bs{d}_n(t)$ along the nuclear spin orientation $\hat{\bs{\sigma}}_n$ \cite{graham2013new},
\begin{align}
\bs{d}_n(t) = g_d a(\bs{r},t) \hat{\bs{\sigma}}_n~,
\label{eq:oscillating-EDM}
\end{align}
where $g_d$ is the coupling parameter (inversely proportional to the associated symmetry-breaking scale $f_a$).
Axions can also couple directly to Standard Model spins $\hat{\bs{\sigma}}$ through the gradient interaction \cite{graham2013new}, described for nuclear spins by the Hamiltonian
\begin{align}
\mc{H}_g = g_{aNN} \bs{\nabla} a(\bs{r},t) \cdot \hat{\bs{\sigma}}_n~,
\label{eq:gradient-interaction}
\end{align}
which, in analogy with the Zeeman effect, shows that $\bs{\nabla} a(\bs{r},t)$ acts as a pseudo-magnetic field with amplitude $B_a$:
\begin{align}
B_a \approx \frac{g_{aNN}}{\hbar\gamma_n} \sqrt{ 2\hbar^3v^2c \rho_{dm} }~,
\end{align}
where $\gamma_n$ is the nuclear gyromagnetic ratio. (An analogous situation occurs for other fermions, but characterized by different coupling constants.)

\begin{figure}
\center
\includegraphics[width=14cm]{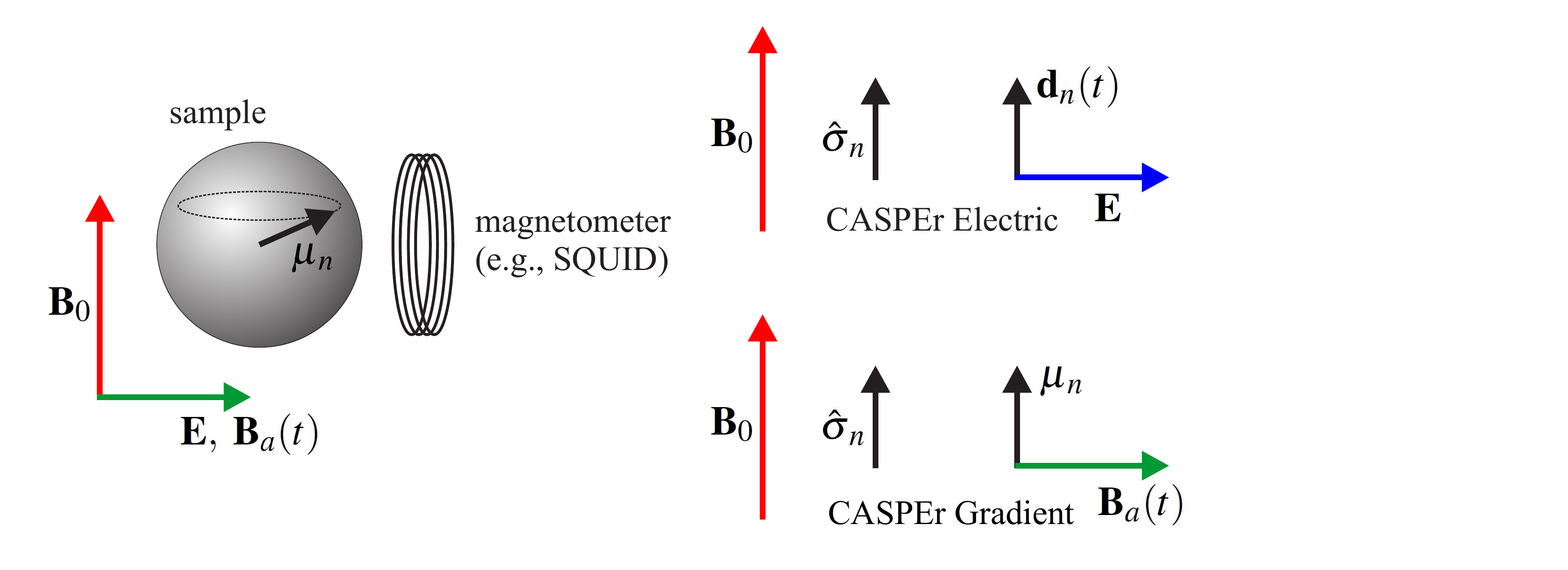}
\caption{{\emph{Left-hand side}}: Schematic diagram of the CASPEr experiment. When the Larmor frequency matches the axion Compton frequency, $\Omega_L \approx \omega_c$, the nuclear spins in the sample are tipped away from their initial orientation along $\bs{B}_0$ due to the axion-induced torque. The precessing magnetization at $\Omega_L$ can be detected with a magnetometer (such as a SQUID) placed near the sample. {\emph{Right-hand side}}: Experimental geometries for CASPEr Electric (top) and CASPEr Gradient (bottom). In both cases, the nuclear spins $\hat{\bs{\sigma}}_n$ are oriented along a leading magnetic field $\bs{B}_0$. An oscillating torque, $\bs{\tau}\ts{EDM} = \bs{d}_n(t) \times \bs{E}$ in the case of CASPEr Electric and $\bs{\tau}\ts{grad} = \bs{\mu}_n \times \bs{B}_a(t)$ in the case of CASPEr Gradient, tips the nuclear spins away from $\bs{B}_0$ if the Larmor frequency $\Omega_L$ matches $\omega_c$. Figure adapted from Ref.~\cite{jackson2020overview}.}
\label{Fig:CASPErSetup}
\end{figure}

In either case, there appears an oscillating torque on spins due to the axion field.
For the axion-gluon (EDM) interaction of Eq.~\eqref{eq:oscillating-EDM} this torque is given by
\begin{align}
\bs{\tau}\ts{EDM} = \bs{d}_n(t) \times \bs{E}^*,
\label{Eq:EDM-torque}
\end{align}
where $\bs{E}^*$ is an effective electric field, which depends on the atomic and nuclear structure of the spin system under study~\cite{Ayb21CASPErE}.
For the axion-fermion interaction of Eq.~\eqref{eq:gradient-interaction} this torque is
\begin{align}
\bs{\tau}\ts{grad} = \bs{\mu}_n \times \bs{B}_a(t),
\label{Eq:gradient-torque}
\end{align}
where $\bs{\mu}_n \propto \hat{\bs{\sigma}}_n$ being the nuclear magnetic moment.
Therefore the interaction between an axion dark matter field and nuclear spins is equivalent to that of an oscillating magnetic field as illustrated in Fig.~\ref{Fig:CASPErSetup}, and consequently the tools of magnetic resonance can be used to search for axion dark matter. 
This is the central concept of the Cosmic Axion Spin Precession Experiment (CASPEr) \cite{budker2014proposal,Garcon2017,wu2019search,garcon2019constraints,Ayb21CASPErE}.

Nuclear magnetic resonance (NMR) experiments involve measuring nuclear spin dynamics in an applied bias field $\bs{B}_0$ that determines the Larmor frequency $\Omega_L = \gamma_n B_0$, although $\bs{B}_0$ can be near zero in zero-to-ultralow field (ZULF) NMR experiments \cite{Led13ZULFNMR} -- a technique used in Refs.~\cite{wu2019search,garcon2019constraints}.
In CASPEr, like other dark matter haloscope experiments, the oscillating field is assumed to always be present, corresponding to case of continuous-wave (cw) NMR.
The magnetic field is scanned, and if $\Omega_L \approx \omega_c$, a resonance occurs and the spins are tilted away from the direction of $\bs{B}_0$ and precess at $\Omega_L$, generating a time-dependent magnetization that can be measured, for example, by induction through a pick-up loop or with a SQUID.

The CASPEr experimental program is divided into two branches: CASPEr Electric, which searches for an oscillating EDM $\bs{d}_n(t)$, and CASPEr Gradient, which searches for an oscillating pseudo-magnetic field $\bs{B}_a(t)$~\cite{jackson2020overview}.
A key to CASPEr's sensitivity is the coherent ``amplification'' of the effects of the axion dark matter field through a large number of polarized nuclear spins.
Therefore an important technological development is the ability to carry out NMR on the largest possible number of spins: this requires large nuclear spin ensembles with high polarization, a focus of CASPEr research efforts, which include thermal polarization, optical polarization, and dynamic nuclear polarization~\cite{Aybas2021}.
Another area of focus is optimization of spin ensemble coherence time, making use of quantum control and decoupling schemes~\cite{Aybas2021}. 
Identifying the optimal spin species and materials with large effective electric fields is especially important for CASPEr Electric, where the detectable signal is proportional to $E^*$. Optimal atomic systems are heavy (large atomic number $Z$) and optimal materials have broken inversion symmetry, such as ferroelectric solids~\cite{Ayb21CASPErE}.
Optimizing the coupling of the spin ensemble to the readout sensor that measures its dynamics is yet another area of focus. Quantum back-action effects will eventually limit the sensitivity of NMR experiments to axion dark matter, and therefore back-action evasion techniques will need to be developed for CASPEr experiments approaching fundamental spin projection noise sensitivity limits~\cite{Aybas2021}.

The QUAX (QUaerere AXion) experiment \cite{barbieri2017searching,crescini2018operation,crescini2020axion} searches for axion dark matter in a manner similar to CASPEr but by exploiting the interaction of axions with electron spins.
The QUAX experiment searches for a coupling of the form \eqref{eq:gradient-interaction} but with the nuclear coupling $g_{aNN}$ replaced by the electron coupling $g_{aee}$, and the electron spin $\hat{\bs{\sigma}}_e$ playing the role of the nuclear spin $\hat{\bs{\sigma}}_n$. 
Ten spherical yttrium iron garnet (YIG) samples are coupled to a cylindrical copper cavity by means of an applied static magnetic field, and the resulting photon-magnon hybrid system acts as an axion-to-electromagnetic field transducer.
This transducer is then coupled to a sensitive rf detector (a quantum-limited Josephson parametric amplifier).
The QUAX experiment is one of the most sensitive rf spin magnetometers ever realized, able to measure fields as small as $5.5 \times 10^{-19}$~T with nine hours of integration time \cite{crescini2020axion}.

Clearly, there is significant overlap between CASPEr and QUAX techniques and those used to search for static EDMs (Sec.~\ref{sec:EDMs}) and exotic spin-dependent interactions (Sec.~\ref{sec:MagsComags}). 
Indeed, in Refs.~\cite{bloch2020axion,jiang2021search,bloch2022new,terrano2019constraints,abel2017search,roussy2021experimental}, noble gas comagnetometers, a spin-polarized torsion pendulum, and apparatuses used for EDM experiments were used as spin-based haloscopes to place limits on axionlike dark matter in the low mass range, corresponding to low Compton frequencies.
Of note are the development of Floquet masers \cite{jiang2021floquet} and spin-amplifiers \cite{jiang2021floquet2} that may expand the nominal bandwidth of noble gas comagnetometers and enable parallel dark matter searches in different frequency ranges.

The Axion Resonant InterAction Detection Experiment (ARIADNE) experiment \cite{arvanitaki2014resonantly,progressariadne} is another example of how spin-based sensors can be employed to search for new physics.
ARIADNE, like CASPEr and QUAX, aims to use magnetic resonance techniques to search for axions and ALPs, and specifically targets the QCD axion.
ARIADNE employs an unpolarized source mass and a spin-polarized $^3$He low-temperature gas to search for a QCD-axion-mediated spin-dependent interaction: the monopole-dipole coupling described by Eq.~\eqref{Eq:Vps} and discussed in Sec.~\ref{sec:MagsComags}.
In contrast to dark matter haloscopes like CASPEr and QUAX, whose signals depend on the local dark matter density at the Earth, the signal in the ARIADNE experiment does not require axions to constitute dark matter and can be modulated in a controlled way. 
ARIADNE probes QCD axion masses in the higher end of the traditionally allowed axion window, up to $6$~meV, a mass range inaccessible to any other existing experiment. Thus ARIADNE fills an important gap in the search for the QCD axion in this important region of parameter space.

For the QCD axion, the scalar and dipole coupling constants $g_s^{(N)}$ and $g_p^{(N)}$ appearing in Eq.~\eqref{Eq:Vps} are correlated with the axion mass $m$.
As discussed earlier, the axion-mediated spin-dependent interaction manifests as a pseudo-magnetic field $\bs{B}_a$ and can be used to resonantly drive spin precession in the laser-polarized cold $^3$He gas.
This is accomplished by spinning an unpolarized tungsten mass sprocket near the $^3$He vessel. 
As the teeth of the sprocket pass by the sample at $\Omega_L$, the magnetization in the longitudinally polarized He gas begins to precess about the axis of an applied field.
This precessing transverse magnetization is detected with a SQUID).
The $^3$He sample acts as an amplifier to transduce the small fictitious magnetic field $B_a$ into a larger real magnetic field detectable by the SQUID, similar to the approach of the CASPEr Gradient experiment \cite{jackson2020overview}.
Superconducting shielding is needed around the sample to screen it from ordinary magnetic field noise which would otherwise limit the sensitivity of the measurement. 
The ARIADNE experiment sources the axion field in the lab, and can explore all mass ranges in the sensitivity band simultaneously, unlike other haloscope experiments which must scan over the possible axion oscillation frequencies $\omega_c$ by tuning a magnetic field \cite{budker2014proposal,Ayb21CASPErE} or cavity \cite{du2018search,zhong2018results}.

Future prospects for improvements in the search for novel spin dependent interactions could include investigations with a spin polarized source mass, or improved sensitivity with new cryogenic or quantum technologies.
Spin squeezing or coherent collective modes in $^3$He could offer prospects for improved sensitivity beyond the Standard quantum limit of spin projection noise in experiments such as ARIADNE, potentially allowing sensitivity all the way down to the SQUID-limited sensitivity.
This would allow one to rule out the axion over a wide range of masses, and when combined with other promising techniques \cite{budker2014proposal,Ayb21CASPErE,gramolin2021search,ouellet2019first,godfrey2021search}, and existing experiments \cite{du2018search,zhong2018results} already at QCD axion sensitivity, could allow in principle the QCD axion to be searched for over its entire allowed mass range.

\section{Spin-based sensors as dark matter particle detectors}
\label{sec:defects}

The scattering of dark matter in crystals is a well developed approach to search for canonical weakly interacting massive particle (or WIMP) dark matter. Searches for WIMP dark matter are soon expected to hit an irreducible background, namely, the coherent scattering of neutrinos from the Sun. This problem is particularly acute for low mass WIMPs with the mass of a few GeV. There are important scientific reasons to probe WIMP cross-sections below the neutrino floor since such cross-sections are natural in models where the WIMP interacts with the Standard Model via the Higgs boson. One way to probe the dark matter parameter space below the neutrino floor is to develop detectors that are able to identify the direction of the nuclear recoil caused by the scattering of dark matter. Since the location of the Sun is known, one may veto all scattering events that point away from the Sun, rejecting all events due to solar neutrinos. The dark matter, being isotropic, will induce scattering events in all directions, permitting an unambiguous detection. The key challenge that needs to be overcome to implement this concept is that directional detection needs to be accomplished in a sample with a large enough (ton scale) target mass since the WIMP cross-sections of interest are tiny. For a practical detector, this requires the ability to perform directional detection in the solid/liquid state so that the detector is sufficiently compact. 

This challenge could conceivably be met in a solid state detector via the following concept \cite{Rajendran:2017ynw}. The scattering of the dark matter displaces an atom off its lattice location and the displaced atom kicks many other atoms off their locations. This causes a tell tale damage track, $\sim 10 - 100$ nm, in the crystal that points to the direction of the incoming dark matter. The created damage can be measured using techniques established in the fields of solid-state quantum sensing and quantum information processing. The detection concept would utilize conventional localization techniques to identify the location of an event of interest to within $\sim$ mm precision. Diffraction limited optics can then be used to achieve micron scale localization. Optical superresolution or high resolution X-ray nanoscopy techniques can then be used to measure the damage track at the nanometer scale. One way to accomplish this superresolution imaging is to use NV center spin spectroscopy in polycrystalline diamond.  This technique can also be implemented in a variety of other wide bandgap semiconductors such as divacancies in silicon carbide. 

In the near term, work towards such a solid-state, WIMP detector with directional sensitivity is centered around demonstrating the capability to locate and determine the direction of nuclear recoil damage tracks in diamond or other crystals. This requires adaptation and development of existing techniques, but the current state of the art is not far from the requisite sensitivity and resolution. In the medium term, such a detector will require position-sensitive instrumentation with spatial resolution at the millimeter scale, as well as development of crystal growth techniques to create large volumes of radiopure, structurally homogeneous crystals. With appropriate development, this approach offers a viable path towards directional WIMP detection with sensitivity below the neutrino limit.

Spin-based sensors may also be useful as low-mass dark matter particle detectors. 
For low-mass dark matter particles, not only are interactions rare because of the exceedingly small cross-sections but also the deposited energy in the detector is extremely small, so both high sensitivity and low background are required.
In Ref.~\cite{lyon2022single}, a new method for detecting low-mass dark matter particles is proposed.
The idea is that if a dark matter particle deposits a small amount of energy ($\gtrsim 1~{\rm{meV}}$) into a high-quality crystalline solid, that energy will eventually be converted into ballistic phonons travelling to the crystal surface.
If the crystal surface is covered by a van der Waals liquid helium film, the phonons can cause quantum evaporation of He atoms.
At low temperature (below $\sim 100$~mK) $^3$He atoms in liquid helium reside at the surface in Andreev bound states \cite{andreev1966surface}. 
After being evaporated, the $^3$He atoms can be collected on another surface covered with a van der Waals film of isotopically enriched $^4$He. 
The $^3$He atoms can be localized at mK temperatures to bound electon states on this second helium film \cite{williams1971deformation}, and subsequently detected by sensing their magnetic moments, by measuring, for example, decoherence of electron spin qubits \cite{lyon2006spin}.
This methodology opens the possibility of single $^3$He atom detection and dark matter particle detection at the $\sim 1$~meV scale \cite{lyon2022single}.

\section{Conclusion}
\label{sec:conclusion}

Much of what is now known about the structure and composition of molecules and materials was originally revealed through spin-based measurements such as nuclear magnetic resonance and electron spin resonance. As quantum information science (QIS) continues to advance the level of control over spin systems, new opportunities are emerging to use the same techniques to search for new fundamental physics in a parallel and complementary manner to large scale particle accelerators and direct particle detectors. There are a range of spin-based experiments that can be employed to search for a variety of effects. Searches for permanent electric dipole moments with atoms, molecules, and spins in solids can search for symmetry violations and thereby test possible explanations for the matter-antimatter imbalance in the universe. Spin-based magnetometers and global networks of such detectors can search for and constrain new particles and fields. Spins in solids can also serve as novel particle detectors by using them as \textit{in-situ} probes for the signatures left behind from particle impacts, and $^3$He spins evaporated from liquid helium films on crystal surfaces could be used as low-mass dark matter particle detectors.  While many such efforts are already underway, there remain tremendous opportunities for innovations in spin-based quantum sensors that will enhance their sensitivity, accuracy, and range of potential fundamental physics targets.

\newpage

\Acknowledgements{The authors are sincerely grateful for insightful comments and suggestions from Nicol\`{o} Crescini,  Reuben Shuker, and Andrew Jayich.}








\bibliographystyle{JHEP}
\bibliography{main-precision-spin}


\end{document}